\documentclass[12pt,english,aps,superscriptaddress,preprintnumbers]{revtex4-1}
\usepackage[T1]{fontenc}
\usepackage[latin9]{inputenc}
\usepackage{float}
\usepackage{mathrsfs}
\usepackage{amsmath}
\usepackage{amssymb}
\usepackage{esint}

\makeatletter

\usepackage{mathrsfs}

\@ifundefined{textcolor}{}{%
 \definecolor{BLACK}{gray}{0}
 \definecolor{WHITE}{gray}{1}
 \definecolor{RED}{rgb}{1,0,0}
 \definecolor{GREEN}{rgb}{0,1,0}
 \definecolor{BLUE}{rgb}{0,0,1}
 \definecolor{CYAN}{cmyk}{1,0,0,0}
 \definecolor{MAGENTA}{cmyk}{0,1,0,0}
 \definecolor{YELLOW}{cmyk}{0,0,1,0}
}

\usepackage{babel}

\makeatother

\usepackage{babel}
\begin{document}
\title{Thermodynamics for a rotating chiral fermion system in the uniform
magnetic field}
\author{Ren-Hong Fang}
\email{fangrh@sdu.edu.cn}

\affiliation{Key Laboratory of Particle Physics and Particle Irradiation (MOE),
Institute of Frontier and Interdisciplinary Science, Shandong University,
Qingdao, Shandong 266237, China}
\affiliation{Theoretical Physics Research and Innovation Team, College of Intelligent
Systems Science and Engineering, Hubei Minzu University, Enshi, Hubei
445000, China}
\begin{abstract}
We study the thermodynamics for a uniformly rotating system of chiral
fermions under the uniform magnetic field. Then we obtain the mathematical
expressions of some thermodynamic quantities in terms of the series
with respect to the external magnetic field $B$, the angular velocity
$\Omega$ and the chemical potential $\mu$, expanded around $B=0$,
$\Omega=0$ and $\mu=0$. Our results given by such series are a generalization
of the expressions available in the references simply corresponding
to the lower-order terms of our findings. The zero-temperature limit
of our results is also discussed.
\end{abstract}
\maketitle

\section{Introduction}

The properties of Dirac fermion system have been investigated from
many aspects for a long time. For a hydrodynamic system consisting
of Dirac fermions under the background of electromagnetic fields,
Wigner function is an appropriate tool, which can provide a covariant
and gauge invariant formalism \citep{Elze:1986qd,Vasak:1987um}. It
is worth pointing out that, although Wigner function defined in 8-dimensional
phase space is not always non-negative, one can always obtain non-negative
probability density when the 4-dimensional momentum is integrated
out. For massless (or chiral) fermion system with uniform vorticity
and electromagnetic fields, the charge current and the energy-momentum
tensor up to the second order have been obtained from Wigner function
approach, including chiral anomaly equation, chiral magnetic and vortical
effects \citep{Gao:2012ix,Yang:2020mtz}. The pair production in parallel
electric and magnetic fields with finite temperature and chemical
potential from Wigner function approach is also investigated recently
\citep{Sheng:2018jwf}. Without external electromagnetic fields, the
energy-momentum tensor and charge current of the massless fermion
system up to second order in vorticity have been obtained from thermal
field theory \citep{Buzzegoli:2017cqy,Buzzegoli:2018wpy,Palermo:2021hlf}.
For a uniformly rotating massless fermion system, the analytic expressions
of the charge current and the energy-momentum tensor are obtained
\citep{Ambrus:2014uqa}. For the massive and massless fermion systems
under the background of a uniform magnetic field, the general expansions
with respect to fermion mass, magnetic field and chemical potential
are derived by the approaches of proper-time and grand partition function
\citep{Cangemi:1996tp,Zhang:2020ben,Fang:2021ndj}. There are also
some investigations on the system of free fermion gas, quark matter
or hadronic matter, with pure rotation \citep{Chen:2021aiq,Fujimoto:2021xix,Becattini:2021lfq},
or with the coexistence of rotation and magnetic field \citep{Chen:2015hfc,Fukushima:2020ncb},
or with specific boundary conditions \citep{Chernodub:2016kxh,Chernodub:2017mvp,Chernodub:2017ref,Zhang:2020hct,Yang:2021hor}.
The quantum superfluid phenomena of Dirac fermions in the background
of magnetic field and rotation are discussed recently \citep{Liu:2017zhl,Mottola:2019nui}. 

In this article, we consider a uniformly rotating chiral fermion system
in a uniform magnetic field, where we ignore the interaction among
the fermions and the directions of the angular velocity and the magnetic
field are chosen to be parallel. In this article we will adopt the
approach of normal ordering and ensemble average to calculate the
thermodynamical quantities of the system. Firstly we briefly derive
the Dirac equation in a rotating frame under the background of a uniform
magnetic field from the Dirac equation in curved space. Then through
solving the eigenvalue equation of the Hamiltonian in cylindrical
coordinates, we can obtain a series of Landau levels, from which one
can calculate the expectation value of corresponding thermodynamical
quantities for each eigenstate. From the approach of ensemble average
used in \citep{Vilenkin:1978hb,Vilenkin:1979ui,Vilenkin:1980fu},
the macroscopic thermodynamical quantities can be expressed as the
summation over the product of the particle number (Fermi-Dirac distribution)
and the expectation value in each eigenstate. We expand all thermodynamical
quantities as threefold series at $B=0$, $\Omega=0$ and $\mu=0$,
where the lower orders are consistent with that from the approaches
of thermal field theory and Wigner function respectively \citep{Buzzegoli:2017cqy,Buzzegoli:2018wpy,Yang:2020mtz},
and to our knowledge the general orders have not been obtained before.
We also calculate all quantities in zero temperature limit, and obtain
the equality of partcile/energy density and corresponding currents
along $z$-axis, which can provide qualitative reference for the thermodynamics
of compact stars in astrophysics, such as neutron star and magnetic
star, since the magnitudes of the magnetic field and rotational speed
are huge compared to the temperature of the compact stars \citep{Felipe:2007vb,Itokazu:2018lij}.
In this article all thermodynamical quantities will be calculated
at the rotating axis ($r=0$), so the boundary condition at the speed-of-light
surface will not affect our results.

From the point of view of hydrodynamics, it has been pointed out that
the relativistic hydrodynamical equations with only first order term
does not obey the causality \citep{Hiscock:1983zz,Hiscock:1985zz,Hiscock:1987zz},
i.e., the group speed of some transport coefficients, such as heat
conductivity, would exceed the speed of light \citep{Denicol:2008ha}.
Therefore, the high order terms in hydrodynamics are necessary, which
indeed repair the issue of causality. There have been some earlier
work to study the second order terms of transport phenomena, such
as Kubo formula from quantum field theory \citep{Jimenez-Alba:2015bia,Hattori:2016njk},
thermal field theory \citep{Buzzegoli:2017cqy,Buzzegoli:2018wpy},
Wigner function \citep{Yang:2020mtz,Yang:2022ksq}, etc. All of these
work are perturbation theory essentially, from which the general order
terms have not been obtained. In this article, we consider a special
configuration for the electromagnetic field and vorticity field in
hydrodynamics, i.e. with a pure homogeneous magnetic field parallel
to a homogeneous vorticity field, and obtain the general order terms
of all thermodynamic quantities, which is important to study the analytic
behavior of hydrodynamics in mathematics.

The rest of this article is organized as follows. In Sec. \ref{sec:2}
and \ref{sec:3}, we briefly derive the Dirac equation in a uniformly
rotating frame and list the Landau levels and corresponding eigenfunctions
of a single right-handed fermion, which are just reference review.
In Sec. \ref{sec:4} and \ref{sec:5}, we obtain the expressions of
some thermodynamic quantities in terms of the series with respect
to the external magnetic field $B$, the angular velocity $\Omega$
and the chemical potential $\mu$, expanded around $B=0$, $\Omega=0$
and $\mu=0$, which are our main results. In Sec. \ref{sec:6}, the
zero temperature limit of the thermodynamical quantities is discussed.
This article is summarized in Sec. \ref{sec:7}.

Throughout this article we adopt natural units where $\hbar=c=k_{B}=1$.
We use the Heaviside-Lorentz convention for electromagnetism and the
chiral representation for gamma matrixes where $\gamma^{5}=\mathrm{diag\,(-1,-1,+1,+1)}$,
which is the same as Peskin and Schroeder \citep{Peskin:1995}. 

\section{Dirac equation in a uniformly rotating frame}

\label{sec:2}In this section we briefly introduce the Dirac equation
in curved spacetime \citep{Parker:2009}, which is applied to a uniformly
rotating frame \citep{Chen:2015hfc}. 

In curved spacetime, under the background of the electromagnetic field,
the Dirac equation for a single chiral fermion is 
\begin{equation}
i\underline{\gamma}^{\mu}D_{\mu}\psi(x)=0,\label{eq:g1}
\end{equation}
where the covariant derivative $D_{\mu}$ and gamma matrices $\underline{\gamma}^{\mu}$
are defined as
\begin{equation}
D_{\mu}=\partial_{\mu}+ieA_{\mu}+\Gamma_{\mu},\ \ \ \underline{\gamma}^{\mu}=\gamma^{a}e_{a}^{\ \mu}.\label{eq:g2}
\end{equation}
The underline in $\underline{\gamma}^{\mu}$ is used to distinguish
the spacetime-dependent gamma matrices $\underline{\gamma}^{\mu}$
from the constant gamma matrices $\gamma^{a}$, and $\Gamma_{\mu}=\frac{1}{8}\omega_{\mu ab}[\gamma^{a},\gamma^{b}]$
is the affine connection. The definitions of vierbein $e_{a}^{\ \mu}$,
metric tensor $g_{\mu\nu}$, and spin connection $\omega_{\mu ab}$
are listed as follows,
\begin{equation}
e_{a}^{\ \mu}=\frac{\partial x^{\mu}}{\partial X^{a}},\ \ \ e_{\ \mu}^{a}=\frac{\partial X^{a}}{\partial x^{\mu}},\ \ \ g_{\mu\nu}=\eta_{ab}e_{\ \mu}^{a}e_{\ \nu}^{b},\label{eq:a2}
\end{equation}
\begin{equation}
\omega_{\mu ab}=g_{\alpha\beta}e_{a}^{\ \alpha}(\partial_{\mu}e_{b}^{\ \beta}+\Gamma_{\mu\nu}^{\beta}e_{b}^{\ \nu}),\label{eq:g3}
\end{equation}
\begin{equation}
\Gamma_{\mu\nu}^{\beta}=\frac{1}{2}g^{\beta\sigma}(g_{\sigma\mu,\nu}+g_{\sigma\nu,\mu}-g_{\mu\nu,\sigma}),\label{eq:g4}
\end{equation}
where $\eta_{ab}=\mathrm{diag}\,(+1,-1,-1,-1)$ is the metric tensor
in Minkowski space, $X^{a}$ and $x^{\mu}$ are the coordinates in
a local Lorentz frame and in a general frame, respectively. 

In curved spacetime, the vector $J_{V}^{\mu}$, axial vector $J_{A}^{\mu}$
and symmetric energy-momentum tensor $T^{\mu\nu}$ become
\begin{equation}
J_{V}^{\mu}=\overline{\psi}\underline{\gamma}^{\mu}\psi,\ \ \ \ J_{A}^{\mu}=\overline{\psi}\underline{\gamma}^{\mu}\gamma^{5}\psi,\label{eq:c3}
\end{equation}
\begin{equation}
T^{\mu\nu}=\frac{1}{4}\left(\overline{\psi}i\underline{\gamma}^{\mu}D^{\nu}\psi+\overline{\psi}i\underline{\gamma}^{\nu}D^{\mu}\psi+\mathrm{H.C.}\right),\label{eq:c4}
\end{equation}
where $D^{\mu}$, $\underline{\gamma}^{\mu}$ in curved spacetime
have replaced $\partial^{a}$, $\gamma^{a}$ in flat spacetime.

Now we consider a frame $\mathcal{K}$ rotating uniformly with angular
velocity $\boldsymbol{\Omega}=\Omega\mathbf{e}_{z}$ relative to an
inertial frame $K$. The coordinates in $\mathcal{K}$ and $K$ are
denoted as $x^{\mu}=(t,x,y,z)$ and $X^{a}=(T,X,Y,Z)$ respectively,
which are related to each other by following transformations,
\begin{equation}
\left\{ \begin{array}{ccc}
T & = & t\hspace{3cm}\\
X & = & x\cos\Omega t-y\sin\Omega t\\
Y & = & x\sin\Omega t+y\cos\Omega t\\
Z & = & z\hspace{3cm}
\end{array}\right..\label{eq:g5}
\end{equation}
It should be pointed out that the rotational angular velocity $\Omega$
can not be too large, otherwise the synchronous condition in Eq. (\ref{eq:g5})
can not be satisfied. According to Eq. (\ref{eq:a2}), the metric
tensor $g_{\mu\nu}$ and its inverse are
\begin{equation}
g_{\mu\nu}=\left(\begin{array}{cccc}
1-(x^{2}+y^{2})\Omega^{2} & y\Omega & -x\Omega & 0\\
y\Omega & -1 & 0 & 0\\
-x\Omega & 0 & -1 & 0\\
0 & 0 & 0 & -1
\end{array}\right),\label{eq:g6}
\end{equation}
\begin{equation}
g^{\mu\nu}=\left(\begin{array}{cccc}
1 & y\Omega & -x\Omega & 0\\
y\Omega & y^{2}\Omega^{2}-1 & -xy\Omega^{2} & 0\\
-x\Omega & -xy\Omega^{2} & x^{2}\Omega^{2}-1 & 0\\
0 & 0 & 0 & -1
\end{array}\right).\label{eq:g7}
\end{equation}
Keeping $g_{\mu\nu}$ unchanged, the vierbein $e_{\ \mu}^{a}$ still
has a freedom degree of an arbitrary local Lorentz transformation.
We can choose $e_{\ \mu}^{a}$ as
\begin{equation}
e_{\ 0}^{0}=e_{\ 1}^{1}=e_{\ 2}^{2}=e_{\ 3}^{3}=1,\ \ \ \ e_{\ 0}^{1}=-y\Omega,\ \ \ \ e_{\ 0}^{2}=x\Omega,\label{eq:g8}
\end{equation}
with zeros for other components.

Now we consider a single chiral fermion in a uniformly rotating frame
under the background of a uniform magnetic field $\mathbf{B}=B\mathbf{e}_{z}$,
and we choose the gauge potential in the inertial frame as $A^{a}=(0,\mathbf{A})$
with $\mathbf{B}=\nabla\times\mathbf{A}$. The covariant derivative
$D_{\mu}$ and gamma matrices $\underline{\gamma}^{\mu}$ become
\begin{equation}
D_{\mu}=\left(\partial_{t}-\frac{i}{2}\Omega\Sigma_{3}+\Omega(yA_{x}-xA_{y}),\partial_{x}-ieA_{x},\partial_{y}-ieA_{y},\partial_{z}-ieA_{z}\right),\label{eq:c1}
\end{equation}
\begin{equation}
\underline{\gamma}^{\mu}=\left(\gamma^{0},y\Omega\gamma^{0}+\gamma^{1},-x\Omega\gamma^{0}+\gamma^{2},\gamma^{3}\right),\label{eq:c2}
\end{equation}
and in this case the Dirac equation for a single chiral fermion can
be written as
\begin{equation}
i\frac{\partial}{\partial t}\psi(x)=\left[-i\gamma^{0}\boldsymbol{\gamma}\cdot(\nabla-ie\mathbf{A})-\Omega J_{z}\right]\psi(x),\label{eq:a3}
\end{equation}
where $e$ is the charge of the chiral fermion, $J_{z}=\frac{1}{2}\Sigma_{3}-i(x\partial_{y}-y\partial_{x})$
is the $z$-component of the total angular momentum $\mathbf{J}$,
and the term $-\Omega J_{z}$ can be naturally explained as the coupling
of the angular momentum $\mathbf{J}$ and the angular velocity $\boldsymbol{\Omega}$. 

\section{Landau levels for a single right-handed fermion in a rotating frame}

\label{sec:3}In the chiral representation of gamma matrixes, where
$\gamma^{5}=\mathrm{diag\,(-1,-1,+1,+1)}$, we can divide the chiral
fermion field into left- and right-handed fermion fields, i.e. $\psi=(\psi_{L},\psi_{R})^{T}$.
Since the equations of motion for $\psi_{L}$ and $\psi_{R}$ decouple,
we only discuss right-handed fermion field in this article. All results
can be directly generalised to the left-handed case. In the following,
we set $eB>0$ for simplicity. 

The right-handed part of Eq. (\ref{eq:a3}) is
\begin{equation}
i\frac{\partial}{\partial t}\psi_{R}(x)=H\psi_{R}(x),\label{eq:g9}
\end{equation}
\begin{equation}
H=-i\boldsymbol{\sigma}\cdot(\nabla-ie\mathbf{A})-\Omega J_{R,z},\label{eq:g10}
\end{equation}
where $H$, $J_{R,z}=\frac{1}{2}\sigma_{3}-i(x\partial_{y}-y\partial_{x})$
is Hamiltonian and the $z$-component of the total angular momentum
of the right-handed fermion. In this article we shall choose the symmetric
gauge for $\mathbf{A}$, i.e. $\mathbf{A}=(-\frac{1}{2}By,\frac{1}{2}Bx,0)$.
Then the explicit form of the Hamiltonian is
\begin{equation}
H=-i\boldsymbol{\sigma}\cdot\nabla+\frac{1}{2}eB\left(y\sigma_{1}-x\sigma_{2}\right)-\Omega J_{R,z}.\label{eq:g11}
\end{equation}
It can be proved that, these three Hermitian operators, $H,\hat{p}_{z}=-i\partial_{z},J_{R,z}$,
are commutative with each other, then we can construct the common
eigenfunctions of them. According to the calculations for Landau levels
in Appendix \ref{sec:app1}, we list the common eigenfunctions and
corresponding energy in cylindrical coordinate system (where the three
coordinate variables are $z,r,\phi$) as follows:

When $m=\frac{1}{2},\frac{3}{2},\frac{5}{2},\cdots$,
\begin{equation}
\psi_{\lambda nmp_{z}}=\sqrt{\frac{n!}{(n+m-\frac{1}{2})!}}\left(\begin{array}{c}
\sqrt{\frac{eB(E+p_{z}+m\Omega)}{2(E+m\Omega)}}e^{-\frac{\rho}{2}}\rho^{\frac{m}{2}-\frac{1}{4}}L_{n}^{m-\frac{1}{2}}(\rho)e^{i(m-\frac{1}{2})\phi}\\
\frac{i\lambda eB}{\sqrt{(E+m\Omega)(E+p_{z}+m\Omega)}}e^{-\frac{\rho}{2}}\rho^{\frac{m}{2}+\frac{1}{4}}L_{n-1}^{m+\frac{1}{2}}(\rho)e^{i(m+\frac{1}{2})\phi}
\end{array}\right)\frac{e^{-iEt+izp_{z}}}{2\pi},\label{eq:b27-1}
\end{equation}
\begin{equation}
E=\left\{ \begin{array}{cc}
\lambda\sqrt{p_{z}^{2}+2eBn}-m\Omega, & n>0\\
p_{z}-m\Omega, & n=0
\end{array}\right..\label{eq:b28-1}
\end{equation}

When $m=-\frac{1}{2},-\frac{3}{2},-\frac{5}{2},\cdots$, 
\begin{equation}
\psi_{\lambda nmp_{z}}=\sqrt{\frac{n!}{(n-m+\frac{1}{2})!}}\left(\begin{array}{c}
\sqrt{\frac{eB(E+p_{z}+m\Omega)}{2(E+m\Omega)}}e^{-\frac{\rho}{2}}\rho^{\frac{1}{4}-\frac{m}{2}}L_{n}^{\frac{1}{2}-m}(\rho)e^{i(m-\frac{1}{2})\phi}\\
-\frac{i\lambda eB(n-m+\frac{1}{2})}{\sqrt{(E+m\Omega)(E+p_{z}+m\Omega)}}e^{-\frac{\rho}{2}}\rho^{-\frac{1}{4}-\frac{m}{2}}L_{n}^{-\frac{1}{2}-m}(\rho)e^{i(m+\frac{1}{2})\phi}
\end{array}\right)\frac{e^{-iEt+izp_{z}}}{2\pi},\label{eq:b25-1}
\end{equation}
\begin{equation}
E=\lambda\sqrt{p_{z}^{2}+2eB\left(n-m+\frac{1}{2}\right)}-m\Omega,\label{eq:b26-1}
\end{equation}
where $\rho=\frac{1}{2}eBr^{2}$, $L_{n}^{\mu}(\rho)$ is general
Laguerre polynomial as introduced in Appendix \ref{sec:app2}, $m$
is the magnetic quantum number, $\lambda=\pm1$ represent the states
with positive and negative energy respectively, and $n=0,1,2,\cdots$
represent different Landau levels. The eigenfunctions $\psi_{\lambda nmp_{z}}$
are denoted by the group of good quantum numbers ($\lambda,n,m,p_{z}$),
which are normalized according to
\begin{equation}
\int dV\psi_{\lambda^{\prime}n^{\prime}m^{\prime}p_{z}^{\prime}}^{\dagger}\psi_{\lambda nmp_{z}}=\delta_{\lambda^{\prime}\lambda}\delta_{n^{\prime}n}\delta_{m^{\prime}m}\delta(p_{z}^{\prime}-p_{z}).\label{eq:g12}
\end{equation}

\section{Particle current}

\label{sec:4}In this section we consider a right-handed fermion system
under the background of a uniform magnetic field $\mathbf{B}=B\mathbf{e}_{z}$,
and the system is rotating uniformly with angular velocity $\boldsymbol{\Omega}=\Omega\mathbf{e}_{z}$.
The interaction among the fermions in this system is ignored. We assume
that this rotating system is in equilibrium with a reservoir, which
keeps constant temperature $T=1/\beta$ and constant chemical potential
$\mu$. 

\subsection{Ensemble average}

We will calculate the macroscopic particle current of the system at
the rotation axis (i.e. at $r=0$) through ensemble average approach,
in which all macroscopic thermodynamical quantities are the ensemble
average of the normal ordering of the corresponding field operators. 

The forms of the eigenfunctions in Eqs. (\ref{eq:b27-1}, \ref{eq:b25-1})
at $r=0$ or $\rho=0$ are simplified to
\begin{equation}
\psi_{\lambda nmp_{z}}=\frac{e^{-iEt+izp_{z}}}{2\pi}\left(\begin{array}{c}
\sqrt{\frac{eB(E+p_{z}+\Omega/2)}{2(E+\Omega/2)}}\delta_{m,1/2}\\
0
\end{array}\right)+\frac{e^{-iEt+izp_{z}}}{2\pi}\left(\begin{array}{c}
0\\
-\frac{i\lambda eB\sqrt{n+1}}{\sqrt{(E-\Omega/2)(E+p_{z}-\Omega/2)}}\delta_{m,-1/2}
\end{array}\right),\label{eq:a9}
\end{equation}
which are to be used in the following calculations of ensemble average.
We find that the $z$-component $m$ of the total angular momentum
can only take values $\pm1/2$ due to the absence of the orbital angular
momentum at $r=0$. 

For the right-handed fermion system, the field operator of the particle
current at $r=0$ is
\begin{equation}
J^{\mu}=\psi_{R}^{\dagger}\sigma^{\mu}\psi_{R},\label{eq:g13}
\end{equation}
with $\sigma^{\mu}=(1,\boldsymbol{\sigma})$. From the approach of
ensemble average used in \citep{Vilenkin:1978hb,Vilenkin:1979ui,Vilenkin:1980fu},
the macroscopic particle current $\mathcal{J}^{\mu}$ can be calculated
from $J^{\mu}$ as follows,
\begin{eqnarray}
\mathcal{J}^{\mu} & = & \langle:J^{\mu}:\rangle\nonumber \\
 & = & \sum_{n=1}^{\infty}\sum_{\lambda}\int_{-\infty}^{\infty}dp_{z}\frac{\lambda}{e^{\beta\left(E_{n}-\lambda\Omega/2-\lambda\mu\right)}+1}\psi_{\lambda,n,1/2,p_{z}}^{\dagger}\sigma^{\mu}\psi_{\lambda,n,1/2,p_{z}}\nonumber \\
 &  & +\sum_{n=0}^{\infty}\sum_{\lambda}\int_{-\infty}^{\infty}dp_{z}\frac{\lambda}{e^{\beta\left(E_{n+1}+\lambda\Omega/2-\lambda\mu\right)}+1}\psi_{\lambda,n,-1/2,p_{z}}^{\dagger}\sigma^{\mu}\psi_{\lambda,n,-1/2,p_{z}}\nonumber \\
 &  & +\sum_{\lambda}\int_{-\infty}^{\infty}dp_{z}\frac{\lambda\theta(\lambda p_{z})}{e^{\beta\left(|p_{z}|-\lambda\Omega/2-\lambda\mu\right)}+1}\psi_{\lambda,0,1/2,p_{z}}^{\dagger}\sigma^{\mu}\psi_{\lambda,0,1/2,p_{z}},\label{eq:f1}
\end{eqnarray}
where $\langle:\cdots:\rangle$ means normal ordering and ensemble
average of corresponding field operator \citep{Vasak:1987um,Dong:2020zci},
$\theta(x)$ is the step function, and we have defined $E_{n}=\sqrt{p_{z}^{2}+2eBn}$.
The second, third, and fourth lines of Eq. (\ref{eq:f1}) represent
the contributions of high Landau levels with $m=1/2$, all Landau
levels with $m=-1/2$, and the lowest Landau level with $m=1/2$,
respectively. We can see that the macroscopic particle current $\mathcal{J}^{\mu}$
consists of the summation over the product of the particle number
(Fermi-Dirac distribution) and the expectation value in each mode
described by the quantum numbers ($\lambda,n,m,p_{z}$).

\subsection{Particle number density}

Firstly we calculate the particle number density $\rho\equiv\mathcal{J}^{0}$
of the system. Making use of 
\begin{equation}
\psi_{\lambda nmp_{z}}^{\dagger}\psi_{\lambda nmp_{z}}=\left\{ \begin{array}{cc}
\frac{eB}{4\pi^{2}}\frac{E+\Omega/2+p_{z}}{E+\Omega/2}, & m=\frac{1}{2}\\
\frac{eB}{4\pi^{2}}\frac{E-\Omega/2-p_{z}}{E-\Omega/2}, & m=-\frac{1}{2}
\end{array}\right.\label{eq:g14}
\end{equation}
and from Eq. (\ref{eq:f1}) one can obtain 
\begin{equation}
\rho\beta^{3}=\frac{b\omega}{16\pi^{2}}+\frac{1}{2}\sum_{s=\pm1}\frac{\partial}{\partial a}g\left(a+\frac{1}{2}s\omega,b\right),\label{eq:e1}
\end{equation}
where we have defined three dimensionless quantities $a=\beta\mu$,
$b=2eB\beta^{2}$, $\omega=\beta\Omega$, and have defined $g(x,b)$
as 
\begin{equation}
g(x,b)=\frac{b}{4\pi^{2}}\int_{0}^{\infty}dy\sum_{n=0}^{\infty}\sum_{s=\pm1}\left(1-\frac{1}{2}\delta_{n,0}\right)\ln\left(1+e^{sx-\sqrt{nb+y^{2}}}\right).\label{eq:g15}
\end{equation}
In a recent article \citep{Zhang:2020ben}, making use of Abel-Plana
formula, the authors obtained the asymptotic expansion of $g(x,b)$
at $b=0$ as follows
\begin{eqnarray}
g(x,b) & = & \left(\frac{7\pi^{2}}{360}+\frac{x^{2}}{12}+\frac{x^{4}}{24\pi^{2}}\right)-\frac{b^{2}\ln b^{2}}{384\pi^{2}}-\frac{b^{2}}{96\pi^{2}}\ln\left(\frac{e}{2G^{6}}\right)\nonumber \\
 &  & -\frac{1}{2\pi^{2}}\sum_{n=0}^{\infty}\frac{(4n+1)!!}{(4n+4)!!}\mathscr{B}_{2n+2}C_{2n+1}(x)b^{2n+2},\label{eq:f3}
\end{eqnarray}
where $G=1.28242...$ is the Glaisher number, $\mathscr{B}_{n}$ are
Bernoulli numbers, and $C_{2n+1}(x)$ is defined and expanded at $x=0$
in the following,
\begin{eqnarray}
C_{2n+1}(x) & = & -\delta_{n,0}+\frac{1}{(4n+1)!}\int_{0}^{\infty}dy\ln y\frac{d^{4n+1}}{dy^{4n+1}}\left(\frac{1}{e^{y+x}+1}+\frac{1}{e^{y-x}+1}\right)\nonumber \\
 & = & (\ln4+\gamma-1)\delta_{n,0}+\frac{2}{(4n+1)!}\sum_{k=0}^{\infty}\left(2^{4n+2k+1}-1\right)\zeta^{\prime}(-4n-2k)\frac{x^{2k}}{(2k)!}.\label{eq:f2}
\end{eqnarray}
Plugging Eqs. (\ref{eq:f3}, \ref{eq:f2}) into Eq. (\ref{eq:e1}),
one can get the threefold series expansion of the particle number
density at $a=0$, $b=0$, $\omega=0$ or $\mu=0$, $B=0$, $\Omega=0$
as follows,
\begin{eqnarray}
\rho\beta^{3} & = & \frac{a}{6}+\frac{a^{3}}{6\pi^{2}}+\frac{a\omega^{2}}{8\pi^{2}}+\frac{b\omega}{16\pi^{2}}\nonumber \\
 &  & -\frac{1}{\pi^{2}}\sum_{n=0}^{\infty}\frac{\mathscr{B}_{2n+2}b^{2n+2}}{(4n+4)!!(4n)!!}\sum_{j=0}^{\infty}\frac{\omega^{2j}}{(2j)!2^{2j}}\nonumber \\
 &  & \times\sum_{k=0}^{\infty}\left(2^{4n+2k+2j+3}-1\right)\zeta^{\prime}(-4n-2k-2j-2)\frac{a^{2k+1}}{(2k+1)!}.\label{eq:d1}
\end{eqnarray}
The lower orders $O(b^{2},\omega^{2},b\omega)$ in Eq. (\ref{eq:d1})
are consistent with the perturbative results in \citep{Buzzegoli:2017cqy,Buzzegoli:2018wpy,Yang:2020mtz},
where the authors used the approaches of thermal field theory and
Wigner function respectively.

\subsection{Particle current along $z$-axis}

Next we calculate the space components of the particle current $\mathcal{J}^{\mu}$.
According to the rotation symmetry along $z$-axis of the system,
the $x$- and $y$-components of $\mathcal{J}^{\mu}$ vanish. The
unique nonzero component is $\mathcal{J}^{z}$. Making use of 
\begin{equation}
\psi_{\lambda nmp_{z}}^{\dagger}\sigma_{3}\psi_{\lambda nmp_{z}}=\left\{ \begin{array}{cc}
\frac{eB}{4\pi^{2}}\frac{E+\Omega/2+p_{z}}{E+\Omega/2}, & m=\frac{1}{2}\\
-\frac{eB}{4\pi^{2}}\frac{E-\Omega/2-p_{z}}{E-\Omega/2}, & m=-\frac{1}{2}
\end{array}\right.\label{eq:g16}
\end{equation}
and from Eq. (\ref{eq:f1}) one can obtain

\begin{equation}
\mathcal{J}^{z}\beta^{3}=\frac{ab}{8\pi^{2}}+\frac{1}{2}\sum_{s=\pm1}s\frac{\partial}{\partial a}g\left(a+\frac{1}{2}s\omega,b\right),\label{eq:a14}
\end{equation}
which can be expanded as the threefold series at $a=0$, $b=0$, $\omega=0$
or $\mu=0$, $B=0$, $\Omega=0$ as follows,

\begin{eqnarray}
\mathcal{J}^{z}\beta^{3} & = & \frac{ab}{8\pi^{2}}+\frac{\omega}{12}+\frac{\omega^{3}}{48\pi^{2}}+\frac{\omega a^{2}}{4\pi^{2}}\nonumber \\
 &  & -\frac{1}{\pi^{2}}\sum_{n=0}^{\infty}\frac{\mathscr{B}_{2n+2}b^{2n+2}}{(4n+4)!!(4n)!!}\sum_{j=0}^{\infty}\frac{\omega^{2j+1}}{(2j+1)!2^{2j+1}}\nonumber \\
 &  & \times\sum_{k=0}^{\infty}\left(2^{4n+2k+2j+3}-1\right)\zeta^{\prime}(-4n-2k-2j-2)\frac{a^{2k}}{(2k)!}.\label{eq:a15}
\end{eqnarray}
When $\omega=0$ or $\Omega=0$ in Eq. (\ref{eq:a15}), one can obtain
$\mathcal{J}^{z}\beta^{3}=\frac{ab}{8\pi^{2}}$, which is the chiral
magnetic effect \citep{Kharzeev:2007jp,Fukushima:2008xe,Son:2009tf,Kharzeev:2010gr,Son:2012wh};
When $b=0$ or $B=0$ and keeping the leading order of $\omega$ in
Eq. (\ref{eq:a15}), one can obtain $\mathcal{J}^{z}\beta^{3}=\frac{\omega}{12}\left(1+\frac{3a^{2}}{\pi^{2}}\right)$,
which is the chiral vortical effect \citep{Landsteiner:2011cp,Golkar:2012kb,Hou:2012xg,Lin:2018aon,Gao:2018jsi,Shitade:2020lfe}. 

\section{Energy-momentum tensor}

\label{sec:5}In this section, we will calculate the energy-momentum
tensor $\mathcal{T}^{\mu\nu}$ (at $r=0$) of the right-handed fermion
system as described in Sec. \ref{sec:4}. According to the rotation
symmetry along $z$-axis, the energy-momentum tensor at $r=0$ are
unchanged under the rotation along $z$-axis, which leads to following
constraints on $\mathcal{T}^{\mu\nu}$:
\begin{equation}
\mathcal{T}^{01}=\mathcal{T}^{02}=\mathcal{T}^{12}=\mathcal{T}^{13}=\mathcal{T}^{23}=0,\ \ \ \mathcal{T}^{11}=\mathcal{T}^{22}.\label{eq:g17}
\end{equation}
The possible nonzero components of $\mathcal{T}^{\mu\nu}$ are $\mathcal{T}^{00}$,
$\mathcal{T}^{11}=\mathcal{T}^{22}$, $\mathcal{T}^{33}$, and $\mathcal{T}^{03}$. 

For the right-handed fermion system, the field operator of the symmetric
energy-momentum tensor at $r=0$ is
\begin{equation}
T^{\mu\nu}=\frac{1}{4}\left(\psi_{R}^{\dagger}i\sigma^{\mu}D_{R}^{\nu}\psi_{R}+\psi_{R}^{\dagger}i\sigma^{\nu}D_{R}^{\mu}\psi_{R}+\mathrm{H.C.}\right),\label{eq:g18}
\end{equation}
with $\sigma^{\mu}=(1,\boldsymbol{\sigma})$ and the right-handed
covariant derivative $D_{R}^{\mu}$ defined as
\begin{equation}
D_{R}^{\mu}=\left(\partial_{t}-\frac{i}{2}\Omega\sigma_{3},-\partial_{x},-\partial_{y},-\partial_{z}\right).\label{eq:g19}
\end{equation}
The macroscopic energy-momentum tensor $\mathcal{T}^{\mu\nu}$ can
be calculated from $T^{\mu\nu}$ as follows,
\begin{eqnarray}
\mathcal{T}^{\mu\nu} & = & \langle:T^{\mu\nu}:\rangle\nonumber \\
 & = & \frac{1}{4}\sum_{n=1}^{\infty}\sum_{\lambda}\int_{-\infty}^{\infty}dp_{z}\frac{\lambda}{e^{\beta\left(E_{n}-\lambda\Omega/2-\lambda\mu\right)}+1}\psi_{\lambda,n,1/2,p_{z}}^{\dagger}\left(i\sigma^{\mu}D_{R}^{\nu}+i\sigma^{\nu}D_{R}^{\mu}\right)\psi_{\lambda,n,1/2,p_{z}}\nonumber \\
 &  & +\frac{1}{4}\sum_{n=0}^{\infty}\sum_{\lambda}\int_{-\infty}^{\infty}dp_{z}\frac{\lambda}{e^{\beta\left(E_{n+1}+\lambda\Omega/2-\lambda\mu\right)}+1}\psi_{\lambda,n,-1/2,p_{z}}^{\dagger}\left(i\sigma^{\mu}D_{R}^{\nu}+i\sigma^{\nu}D_{R}^{\mu}\right)\psi_{\lambda,n,-1/2,p_{z}}\nonumber \\
 &  & +\frac{1}{4}\sum_{\lambda}\int_{-\infty}^{\infty}dp_{z}\frac{\lambda\theta(\lambda p_{z})}{e^{\beta\left(|p_{z}|-\lambda\Omega/2-\lambda\mu\right)}+1}\psi_{\lambda,0,1/2,p_{z}}^{\dagger}\left(i\sigma^{\mu}D_{R}^{\nu}+i\sigma^{\nu}D_{R}^{\mu}\right)\psi_{\lambda,0,1/2,p_{z}}+\mathrm{H.C.}\nonumber \\
\label{eq:f6}
\end{eqnarray}

\subsection{Energy density}

Firstly we calculate the energy density $\varepsilon\equiv\mathcal{T}^{00}$
of the system. Making use of 
\begin{equation}
\psi_{\lambda nmp_{z}}^{\dagger}\left(i\partial_{t}+\frac{1}{2}\Omega\sigma_{3}\right)\psi_{\lambda nmp_{z}}=\left\{ \begin{array}{cc}
\frac{eB}{8\pi^{2}}(E+p_{z}+\Omega/2), & m=\frac{1}{2}\\
\frac{eB}{8\pi^{2}}(E-p_{z}-\Omega/2), & m=-\frac{1}{2}
\end{array}\right.\label{eq:g20}
\end{equation}
and from Eq. (\ref{eq:f1}) one can obtain

\begin{equation}
\varepsilon\beta^{4}=\frac{ab\omega}{16\pi^{2}}+\sum_{s=\pm1}\left(\frac{3}{2}-b\frac{\partial}{\partial b}\right)g\left(a+\frac{1}{2}s\omega,b\right),\label{eq:e2}
\end{equation}
which can be expanded as the threefold series at $a=0$, $b=0$, $\omega=0$
or $\mu=0$, $B=0$, $\Omega=0$ as follows,
\begin{eqnarray}
\varepsilon\beta^{4} & = & \frac{7\pi^{2}}{120}+\frac{a^{2}}{4}+\frac{\omega^{2}}{16}+\frac{a^{4}}{8\pi^{2}}+\frac{3a^{2}\omega^{2}}{16\pi^{2}}+\frac{\omega^{4}}{128\pi^{2}}\nonumber \\
 &  & +\frac{ab\omega}{16\pi^{2}}+\frac{b^{2}\ln b^{2}}{384\pi^{2}}+\frac{b^{2}}{96\pi^{2}}\ln\left(\frac{2e^{\gamma+1}}{G^{6}}\right)\nonumber \\
 &  & +\frac{1}{\pi^{2}}\sum_{n=0}^{\infty}\frac{(4n+1)\mathscr{B}_{2n+2}b^{2n+2}}{(4n+4)!!(4n)!!}\sum_{j=0}^{\infty}\frac{\omega^{2j}}{(2j)!2^{2j}}\nonumber \\
 &  & \times\sum_{k=0}^{\infty}\left(2^{4n+2k+2j+1}-1\right)\zeta^{\prime}(-4n-2k-2j)\frac{a^{2k}}{(2k)!},\label{eq:g21}
\end{eqnarray}
where the logarithmic term $b^{2}\ln b^{2}$ has been discussed in
detail in \citep{Zhang:2020ben}, and its coefficient is independent
of $\omega$ in this work. It is worth noting that there would be
no such logarithmic term if the un-normal ordering description of
field operators was adopted \citep{Sheng:2017lfu,Yang:2020mtz}.

\subsection{Pressure}

The pressure $P$ of the system is $\mathcal{T}^{33}$. Making use
of 
\begin{equation}
\psi_{\lambda nmp_{z}}^{\dagger}\sigma_{3}(-i\partial_{z})\psi_{\lambda nmp_{z}}=\left\{ \begin{array}{cc}
\frac{eB}{8\pi^{2}}\frac{(E+p_{z}+\Omega/2)p_{z}}{E+\Omega/2}, & m=\frac{1}{2}\\
-\frac{eB}{8\pi^{2}}\frac{(E-p_{z}-\Omega/2)p_{z}}{E-\Omega/2}, & m=-\frac{1}{2}
\end{array}\right.\label{eq:g22}
\end{equation}
and from Eq. (\ref{eq:f1}) one can obtain
\begin{equation}
P\beta^{4}=\frac{ab\omega}{16\pi^{2}}+\frac{1}{2}\sum_{s=\pm1}g\left(a+\frac{1}{2}s\omega,b\right),\label{eq:e3}
\end{equation}
which can be expanded as the threefold series at $a=0$, $b=0$, $\omega=0$
or $\mu=0$, $B=0$, $\Omega=0$ as follows,
\begin{eqnarray}
P\beta^{4} & = & \frac{7\pi^{2}}{360}+\frac{a^{2}}{12}+\frac{\omega^{2}}{48}+\frac{a^{4}}{24\pi^{2}}+\frac{a^{2}\omega^{2}}{16\pi^{2}}+\frac{\omega^{4}}{384\pi^{2}}\nonumber \\
 &  & +\frac{ab\omega}{16\pi^{2}}-\frac{b^{2}\ln b^{2}}{384\pi^{2}}-\frac{b^{2}}{96\pi^{2}}\ln\left(\frac{2e^{\gamma}}{G^{6}}\right)\nonumber \\
 &  & -\frac{1}{\pi^{2}}\sum_{n=0}^{\infty}\frac{\mathscr{B}_{2n+2}b^{2n+2}}{(4n+4)!!(4n)!!}\sum_{j=0}^{\infty}\frac{\omega^{2j}}{(2j)!2^{2j}}\nonumber \\
 &  & \times\sum_{k=0}^{\infty}\left(2^{4n+2k+2j+1}-1\right)\zeta^{\prime}(-4n-2k-2j)\frac{a^{2k}}{(2k)!}.\label{eq:g23}
\end{eqnarray}
One can obtain $\mathcal{T}^{11}$ from the traceless condition for
energy-momentum tensor, $\mathcal{T}^{00}=2\mathcal{T}^{11}+\mathcal{T}^{33}$.

\subsection{Energy current}

The energy current along $z$-axis is $\mathcal{T}^{03}$. Making
use of 
\begin{equation}
\psi_{\lambda nmp_{z}}^{\dagger}\left(-i\partial_{z}+\sigma_{3}i\partial_{t}+\frac{1}{2}\Omega\right)\psi_{\lambda nmp_{z}}=\left\{ \begin{array}{cc}
\frac{eB}{8\pi^{2}}\frac{(E+p_{z}+\Omega/2)^{2}}{E+\Omega/2}, & m=\frac{1}{2}\\
-\frac{eB}{8\pi^{2}}\frac{(E-p_{z}-\Omega/2)^{2}}{E-\Omega/2}, & m=-\frac{1}{2}
\end{array}\right.\label{eq:g24}
\end{equation}
and from Eq. (\ref{eq:f1}) one can obtain
\begin{equation}
\mathcal{T}^{03}\beta^{4}=\frac{b}{8\pi^{2}}\left(\frac{\pi^{2}}{6}+\frac{\omega^{2}}{8}+\frac{a^{2}}{2}\right)+\sum_{s=\pm1}s\left(1-\frac{b}{2}\frac{\partial}{\partial b}\right)g\left(a+\frac{1}{2}s\omega,b\right),\label{eq:e4}
\end{equation}
which can be expanded as the threefold series at $a=0$, $b=0$, $\omega=0$
or $\mu=0$, $B=0$, $\Omega=0$ as follows,
\begin{eqnarray}
\mathcal{T}^{03}\beta^{4} & = & \frac{b}{8\pi^{2}}\left(\frac{\pi^{2}}{6}+\frac{\omega^{2}}{8}+\frac{a^{2}}{2}\right)+\left(\frac{a\omega}{6}+\frac{a^{3}\omega}{6\pi^{2}}+\frac{a\omega^{3}}{24\pi^{2}}\right)\nonumber \\
 &  & +\frac{2}{\pi^{2}}\sum_{n=1}^{\infty}\frac{n\mathscr{B}_{2n+2}b^{2n+2}}{(4n+4)!!(4n)!!}\sum_{j=0}^{\infty}\frac{\omega^{2j+1}}{(2j+1)!2^{2j+1}}\nonumber \\
 &  & \times\sum_{k=0}^{\infty}\left(2^{4n+2k+2j+3}-1\right)\zeta^{\prime}(-4n-2k-2j-2)\frac{a^{2k+1}}{(2k+1)!}.\label{eq:g25}
\end{eqnarray}

Up to now, we have obtained all thermodynamical quantities of the
right-handed fermion system. For left-handed fermion system, one can
derive corresponding quantities from the right-handed case through
space inversion: $\rho_{R}\rightarrow\rho_{L}$, $\mathcal{J}_{R}^{z}\rightarrow-\mathcal{J}_{L}^{z}$,
$\varepsilon_{R}\rightarrow\varepsilon_{L}$, $P_{R}\rightarrow P_{L}$,
$\mathcal{T}_{R}^{03}\rightarrow-\mathcal{T}_{L}^{03}$, $\mu_{R}\rightarrow\mu_{L}$,
$B\rightarrow B$, $\Omega\rightarrow\Omega$, where the subscripts
$R,L$ are used to distinguish the quantities in right-handed case
from that in left-handed case.

\section{Zero temperature limit}

\label{sec:6}Now we turn to the thermodynamics of the system at zero
temperature limit. When the temperature tends to be zero, with chemical
potential $\mu$, magnetic field $B$, and angular velocity $\Omega$
fixed, then the three dimensionless quantities $a=\beta\mu$, $b=2eB\beta^{2}$,
$\omega=\beta\Omega$ all tend to be infinity. The asymptotic behavior
of $g(x,b)$ as $x\rightarrow\infty$ and $b\rightarrow\infty$ has
been obtained in \citep{Zhang:2020ben},
\begin{equation}
\lim_{x,b\rightarrow\infty}g(x,b)=\frac{x^{2}b}{16\pi^{2}}.\label{eq:g26}
\end{equation}

From Eqs. (\ref{eq:e1}, \ref{eq:a14}), one can derive the expressions
of the particle density $\rho$ and the current $\mathcal{J}^{z}$
at zero temperature limit as follows,
\begin{equation}
\rho=\mathcal{J}^{z}=\frac{eB}{4\pi^{2}}\left(\mu+\frac{\Omega}{2}\right).\label{eq:g27}
\end{equation}
At zero temperature limit, due to the coupling of the spin with the
magnetic field and the angular velocity, the spin alignment of all
particles and antiparticles will be along $z$-axis of the system.
Since these particles are right-handed, they will move along $z$-axis
with the speed of light $c$ ($c=1$ in natural unit), so it is reasonable
that the particle density $\rho$ equals to the $z$-component current
$\mathcal{J}^{z}$ at zero temperature limit.

From Eqs. (\ref{eq:e2}, \ref{eq:e3}, \ref{eq:e4}), the expressions
of energy density $\varepsilon$, pressure $P$ and energy current
$\mathcal{T}^{03}$ at zero temperature limit are

\begin{equation}
\varepsilon=P=\mathcal{T}^{03}=\frac{eB}{8\pi^{2}}\left(\mu+\frac{\Omega}{2}\right)^{2}.\label{eq:g28}
\end{equation}
The movements of the particles and antiparticles with the speed of
light along $z$-axis leads to the equality of the energy density
$\varepsilon$ and the energy current $\mathcal{T}^{03}$. Since there
is no energy current along the direction of the $x$- and $y$-axis,
then $\mathcal{T}^{11}$ and $\mathcal{T}^{22}$ vanish in this system,
which results in the equality of the energy density $\varepsilon$
and the pressure $P$.

\section{Summary}

\label{sec:7}In this article, we have investigated the thermodynamics
of the uniformly rotating right-handed fermion system under the background
of a uniform magnetic field through the approach of normal ordering
and ensemble average, where all thermodynamical quantities are expanded
as threefold series at $B=0$, $\Omega=0$ and $\mu=0$. For these
threefold series, our results at lower orders are consistent with
previous ones by other authors. It is worth pointing out that the
general orders of $B$ and $\Omega$ in the expressions of the thermodynamical
quantities are obtained for the first time and can provide a useful
reference for the high order calculations from several different approaches,
such as thermal field theory and Wigner function. We also calculate
all quantities in zero temperature limit, and obtain the equality
of partcile/energy density and corresponding currents along $z$-axis.
Since for the chiral fermion the right-handed part decouples from
the left-handed part, in this article we only considered the case
of the right-handed fermion system, which can be directly generalized
to the left-handed case through space inversion. In this article,
the currents and energy-momentum tensor are calculated at the rotating
axis ($r=0$), so the boundary condition at the speed-of-light surface
will not affect our results. The calculations for these quantities
off or far from the rotating axis ($r\ne0$) as well as with the boundary
condition at the speed-of-light surface may be investigated in the
future.

\section{Acknowledgments}

I thank De-Fu Hou for helpful discussion. This work was supported
by the National Natural Science Foundation of China under Grants No.
11890713, and No. 12073008.

\appendix

\section{Landau levels for a single right-handed fermion}

\label{sec:app1}The Hamiltonian for a right-handed fermion under
the background of the uniform magnetic field $\mathbf{B}=B\mathbf{e}_{z}$
is
\begin{equation}
H=-i\boldsymbol{\sigma}\cdot(\nabla-ie\mathbf{A})=-i\boldsymbol{\sigma}\cdot\nabla+\frac{1}{2}eB(y\sigma_{1}-x\sigma_{2}),\label{eq:b8}
\end{equation}
where we have chosen $\mathbf{A}=(-\frac{1}{2}By,\frac{1}{2}Bx,0)$
for the gauge potential. One can refer to \citep{Sheng:2017lfu,Sheng:2019ujr,Dong:2020zci}
for other choices of the gauge potential.

In the following, we will solve the eigenvalue equation of $H$ in
cylindrical coordinates,
\begin{equation}
H\psi=E\psi.\label{eq:a6}
\end{equation}
We can see that the three Hermitian operators, $H$, $\hat{p}_{z}=-i\partial_{z}$,
$\hat{J}_{z}=\frac{1}{2}\sigma_{3}+(x\hat{p}_{y}-y\hat{p}_{x})$ are
commutative with each other, so the eigenfunction $\psi$ can be chosen
as
\begin{equation}
\psi=\left(\begin{array}{c}
f(r)e^{i(m-\frac{1}{2})\phi}\\
ig(r)e^{i(m+\frac{1}{2})\phi}
\end{array}\right)e^{izp_{z}},\label{eq:b9}
\end{equation}
where $-\infty<p_{z}<\infty$ and $m=\pm1/2,\pm3/2,\pm5/2,...$ are
the eigenvalues of $\hat{p}_{z}$ and $\hat{J}_{z}$ respectively.
The explicit form of the Hamiltonian $H$ in cylindrical coordinates
is
\begin{equation}
H=\left(\begin{array}{cc}
-i\frac{\partial}{\partial z} & e^{-i\phi}\left(-i\frac{\partial}{\partial r}-\frac{1}{r}\frac{\partial}{\partial\phi}+\frac{i}{2}eBr\right)\\
e^{i\phi}\left(-i\frac{\partial}{\partial r}+\frac{1}{r}\frac{\partial}{\partial\phi}-\frac{i}{2}eBr\right) & i\frac{\partial}{\partial z}
\end{array}\right),\label{eq:b10}
\end{equation}
then from Eq. (\ref{eq:a6}) we can obtain two differential equations
for $f(r),g(r)$ as follows,
\begin{eqnarray}
(p_{z}-E)f(r)+\left(\frac{\partial}{\partial r}+\frac{m+\frac{1}{2}}{r}-\frac{1}{2}eBr\right)g(r) & = & 0,\label{eq:b11}\\
\left(-\frac{\partial}{\partial r}+\frac{m-\frac{1}{2}}{r}-\frac{1}{2}eBr\right)f(r)+(-p_{z}-E)g(r) & = & 0,\label{eq:a5-1}
\end{eqnarray}
which are equivalent to 
\begin{equation}
\left\{ \frac{\partial^{2}}{\partial r^{2}}+\frac{1}{r}\frac{\partial}{\partial r}-\frac{\left(m-\frac{1}{2}\right)^{2}}{r^{2}}-\left[p_{z}^{2}-E^{2}-eB\left(m+\frac{1}{2}\right)\right]-\frac{1}{4}e^{2}B^{2}r^{2}\right\} f(r)=0\label{eq:a1-1}
\end{equation}
\begin{equation}
g(r)=\frac{1}{p_{z}+E}\left(-\frac{\partial}{\partial r}+\frac{m-\frac{1}{2}}{r}-\frac{1}{2}eBr\right)f(r)\label{eq:a3-2}
\end{equation}
We can define a dimensionless variable $\rho=\frac{1}{2}eBr^{2}$,
then
\begin{equation}
\frac{d}{dr}=eBr\frac{d}{d\rho},\ \ \ \ \frac{d^{2}}{dr^{2}}=eB\frac{d}{d\rho}+2eB\rho\frac{d^{2}}{d\rho^{2}}.\label{eq:a7}
\end{equation}
Now Eq. (\ref{eq:a1-1}) becomes
\begin{equation}
\left\{ \rho\frac{d^{2}}{d\rho^{2}}+\frac{d}{d\rho}-\frac{\left(m-\frac{1}{2}\right)^{2}}{4\rho}-\frac{1}{2eB}\left[p_{z}^{2}-E^{2}-eB\left(m+\frac{1}{2}\right)\right]-\frac{1}{4}\rho\right\} f=0\label{eq:a2-2}
\end{equation}
Next, we choose

\begin{equation}
f=e^{-\frac{\rho}{2}}\rho^{\frac{m}{2}-\frac{1}{4}}G(\rho),\label{eq:b12}
\end{equation}
then Eqs. (\ref{eq:a3-2}, \ref{eq:a2-2}) become
\begin{equation}
g=-\frac{\sqrt{2eB}}{E+p_{z}}e^{-\frac{\rho}{2}}\rho^{\frac{m}{2}+\frac{1}{4}}G^{\prime}(\rho),\label{eq:b13}
\end{equation}
\begin{equation}
\rho G^{\prime\prime}+\left[\left(m+\frac{1}{2}\right)-\rho\right]G^{\prime}-\frac{1}{2eB}\left(p_{z}^{2}-E^{2}\right)G=0.\label{eq:a4-1}
\end{equation}
Define following two quantities,
\begin{equation}
\gamma=m+\frac{1}{2},\ \ \ \ \alpha=\frac{1}{2eB}\left(p_{z}^{2}-E^{2}\right),\label{eq:b14}
\end{equation}
then Eq. (\ref{eq:a4-1}) becomes
\begin{equation}
\rho G^{\prime\prime}+(\gamma-\rho)G^{\prime}-\alpha G=0,\label{eq:b15}
\end{equation}
which is the confluent hypergeometric equation \citep{Zeng:2007}.
With the boundary conditions, $|f(0)|,|f(\infty)|<\infty$, the solutions
for $G(\rho)$, $f(\rho)$, $g(\rho)$ can be chosen as:

(1) When $\gamma=0,-1,-2,...$, i.e. $m=-1/2,-3/2,-5/2$, ..., the
boundary condition $|f(0)|<\infty$ requires that
\begin{equation}
G(\rho)=\rho^{1-\gamma}F(\alpha-\gamma+1,2-\gamma,\rho)=\rho^{\frac{1}{2}-m}F\left(\alpha-m+\frac{1}{2},\frac{3}{2}-m,\rho\right),\label{eq:b16}
\end{equation}
where $F(\alpha,\gamma,\rho)$ is the confluent hypergeometric function
as discussed in Appendix \ref{sec:app2}. In addition, the boundary
condition $|f(\infty)|<\infty$ requires that
\begin{equation}
\alpha-m+\frac{1}{2}=-n\ \ (n\in\mathbb{N}),\ \ \ E=\lambda\sqrt{p_{z}^{2}+2eB\left(n-m+\frac{1}{2}\right)}\ \ \ (\lambda=\pm1),\label{eq:b17}
\end{equation}
\begin{equation}
G(\rho)=\rho^{\frac{1}{2}-m}F\left(-n,\frac{3}{2}-m,\rho\right)\sim\rho^{\frac{1}{2}-m}L_{n}^{\frac{1}{2}-m}(\rho),\label{eq:b18}
\end{equation}
where $L_{n}^{k}(\rho)$ is the general Laguerre polynomial as discussed
in Appendix \ref{sec:app2}. Then one obtain
\begin{equation}
f(\rho)\sim e^{-\frac{\rho}{2}}\rho^{\frac{1}{4}-\frac{m}{2}}L_{n}^{\frac{1}{2}-m}(\rho),\ \ \ \ g(\rho)\sim-\frac{\sqrt{2eB}}{E+p_{z}}\left(n-m+\frac{1}{2}\right)e^{-\frac{\rho}{2}}\rho^{-\frac{1}{4}-\frac{m}{2}}L_{n}^{-\frac{1}{2}-m}(\rho).\label{eq:b19}
\end{equation}

(2) When $\gamma=1,2,3,...$, i.e. $m=1/2,3/2,5/2,...$, the boundary
condition $|f(0)|<\infty$ requires that
\begin{equation}
G(\rho)=F(\alpha,\gamma,\rho)=F\left(\alpha,\frac{1}{2}+m,\rho\right).\label{eq:b20}
\end{equation}
In addition, the boundary condition $|f(\infty)|<\infty$ requires
that
\begin{equation}
\alpha=-n\ \ (n\in\mathbb{N}),\ \ \ E=\lambda\sqrt{p_{z}^{2}+2eBn}\ \ \ (\lambda=\pm1),\label{eq:b21}
\end{equation}
\begin{equation}
G(\rho)=F\left(-n,\frac{1}{2}+m,\rho\right)\sim L_{n}^{m-\frac{1}{2}}(\rho).\label{eq:b22}
\end{equation}
Then one obtain
\begin{equation}
f(\rho)\sim e^{-\frac{\rho}{2}}\rho^{\frac{m}{2}-\frac{1}{4}}L_{n}^{m-\frac{1}{2}}(\rho),\ \ \ \ g(\rho)\sim\frac{\sqrt{2eB}}{E+p_{z}}e^{-\frac{\rho}{2}}\rho^{\frac{m}{2}+\frac{1}{4}}L_{n-1}^{m+\frac{1}{2}}(\rho).\label{eq:b23}
\end{equation}

There is a special case we must point out here: When $m>0$, $n=0$,
we must choose $E=p_{z}$, in which case we have $f(\rho)=e^{-\frac{\rho}{2}}\rho^{\frac{m}{2}-\frac{1}{4}}$,
$g(\rho)=0$. There is no physical solution for $m>0$, $n=0$, $E=-p_{z}$.

Making use of the orthonormal relation of the general Laguerre polynomials,
\begin{equation}
\int_{0}^{\infty}dxe^{-x}x^{\gamma}L_{m}^{\gamma}(x)L_{n}^{\gamma}(x)=\frac{\Gamma(n+\gamma+1)}{n!}\delta_{mn},\label{eq:b24}
\end{equation}
we can obtain the normalized eigenfunctions as follows:

When $m<0$, 
\begin{equation}
\psi_{\lambda nmp_{z}}=\sqrt{\frac{n!}{(n-m+\frac{1}{2})!}}\left(\begin{array}{c}
\sqrt{\frac{eB(E+p_{z})}{2E}}e^{-\frac{\rho}{2}}\rho^{\frac{1}{4}-\frac{m}{2}}L_{n}^{\frac{1}{2}-m}e^{i(m-\frac{1}{2})\phi}\\
-\frac{i\lambda eB(n-m+\frac{1}{2})}{\sqrt{E(E+p_{z})}}e^{-\frac{\rho}{2}}\rho^{-\frac{1}{4}-\frac{m}{2}}L_{n}^{-\frac{1}{2}-m}e^{i(m+\frac{1}{2})\phi}
\end{array}\right)\frac{e^{izp_{z}}}{2\pi},\label{eq:b25}
\end{equation}
\begin{equation}
E=\lambda\sqrt{p_{z}^{2}+2eB\left(n-m+\frac{1}{2}\right)}.\label{eq:b26}
\end{equation}

When $m>0$,
\begin{equation}
\psi_{\lambda nmp_{z}}=\sqrt{\frac{n!}{(n+m-\frac{1}{2})!}}\left(\begin{array}{c}
\sqrt{\frac{eB(E+p_{z})}{2E}}e^{-\frac{\rho}{2}}\rho^{\frac{m}{2}-\frac{1}{4}}L_{n}^{m-\frac{1}{2}}e^{i(m-\frac{1}{2})\phi}\\
\frac{i\lambda eB}{\sqrt{E(E+p_{z})}}e^{-\frac{\rho}{2}}\rho^{\frac{m}{2}+\frac{1}{4}}L_{n-1}^{m+\frac{1}{2}}e^{i(m+\frac{1}{2})\phi}
\end{array}\right)\frac{e^{izp_{z}}}{2\pi},\label{eq:b27}
\end{equation}
\begin{equation}
E=\lambda\sqrt{p_{z}^{2}+2eBn}.\label{eq:b28}
\end{equation}
All normalized eigenfunctions are orthogonal with each other,
\begin{equation}
\int dV\psi_{\lambda^{\prime}n^{\prime}m^{\prime}p_{z}^{\prime}}^{\dagger}\psi_{\lambda nmp_{z}}=\delta_{\lambda^{\prime}\lambda}\delta_{n^{\prime}n}\delta_{m^{\prime}m}\delta(p_{z}^{\prime}-p_{z}).\label{eq:b29}
\end{equation}

\section{Confluent hypergeometric function and Laguerre polynomial}

\label{sec:app2}The confluent hypergeometric equation is \citep{Zeng:2007}
\begin{equation}
zy^{\prime\prime}+(\gamma-z)y^{\prime}-\alpha y=0.\label{eq:b1}
\end{equation}
When $\gamma\notin\mathbb{Z}$, there are two independent solutions
as follows,
\begin{eqnarray}
y_{1} & = & F(\alpha,\gamma,z),\nonumber \\
y_{2} & = & z^{1-\gamma}F(\alpha-\gamma+1,2-\gamma,z),\label{eq:b2}
\end{eqnarray}
where $F(\alpha,\gamma,z)$ is the confluent hypergeometric function
defined as
\begin{equation}
F(\alpha,\gamma,z)=\sum_{k=0}^{\infty}\frac{(\alpha)_{k}}{(\gamma)_{k}}\frac{z^{k}}{k!}\equiv1+\frac{\alpha}{\gamma}z+\frac{\alpha(\alpha+1)}{\gamma(\gamma+1)}\frac{z^{2}}{2!}+\frac{\alpha(\alpha+1)(\alpha+2)}{\gamma(\gamma+1)(\gamma+2)}\frac{z^{3}}{3!}+\cdots.\label{eq:b3}
\end{equation}
The asymptotic behavior of $F(\alpha,\gamma,z)$ as $z\rightarrow\infty$
is the same as $e^{z}$. When $\alpha$ is a non-positive integer,
then $F(\alpha,\gamma,z)$ becomes a polynomial.

The general Laguerre polynomial $L_{n}^{\gamma}(z)$ is defined from
$F(\alpha,\gamma,z)$ as follows \citep{Gradshteyn:2014},
\begin{equation}
L_{n}^{\gamma}(z)=\frac{\Gamma(\gamma+n+1)}{n!\Gamma(\gamma+1)}F(-n,\gamma+1,z)=\left(\begin{array}{c}
\gamma+n\\
n
\end{array}\right)F(-n,\gamma+1,z),\label{eq:b4}
\end{equation}
where $\gamma\in\mathbb{R}$ and $n\in\mathbb{N}$. Laguerre polynomial
$L_{n}^{\gamma}(z)$ satisfies following differential equation
\begin{equation}
zy^{\prime\prime}+(\gamma+1-z)y^{\prime}+ny=0\label{eq:qq21}
\end{equation}
We can rewrite Eq. (\ref{eq:qq21}) as a type of Sturm-Liouvelle equation,
\begin{equation}
\frac{d}{dz}\bigg(z^{\gamma+1}e^{-z}\frac{dy}{dz}\bigg)+nz^{\gamma}e^{-z}y=0,\label{eq:b5}
\end{equation}
which gives the orthogonality of $L_{n}^{\gamma}(z)$,
\begin{equation}
\int_{0}^{\infty}dze^{-z}z^{\gamma}L_{m}^{\gamma}(z)L_{n}^{\gamma}(z)=\frac{\Gamma(n+\gamma+1)}{n!}\delta_{mn}.\label{eq:b6}
\end{equation}
When $\gamma=0$, then $L_{n}^{\gamma}(z)$ becomes the normal Laguerre
polynomial $L_{n}(z)$,
\begin{equation}
L_{n}(z)=L_{n}^{0}(z)=F(-n,1,z).\label{eq:b7}
\end{equation}

\bibliographystyle{apsrev}
\addcontentsline{toc}{section}{\refname}\bibliography{ref-2021-11}

\end{document}